\documentclass[3p,twocolumn]{elsarticle}
\usepackage{graphicx,amssymb}

\usepackage{color}

\def\kforty{$\rm ^{40}K$}
\def\arforty{$\rm ^{40}Ar$}
\def\utwothirtyeight{$\rm ^{238}U$}
\def\thtwothirtytwo{$\rm ^{232}Th$}
\def\rntwotwotwo{$\rm ^{222}Rn$~}
\def\natwentytwo{$\rm ^{22}Na$~}
\def\natwentythree{$\rm ^{23}Na$~}
\def\ionetwentyfive{$\rm ^{125}I$~}

\def\teonetwentyfive{$\rm ^{125}Te$~}

\def\pbtwoten{$\rm ^{210}Pb$}
\def\potwoten{$\rm ^{210}Po$}
\def\bitwoforteen{$\rm ^{214}Bi$}
\def\thtwotwentyeight{$\rm ^{228}Th$}
\def\actwotwentyeight{$\rm ^{228}Ac$}
\def\ratwotwentysix{$\rm ^{226}Ra$}
\def\rntwotwenty{$\rm ^{220}Rn$}
\def\tltwooeight{$\rm ^{208}Tl$}
\def\potwoforteen{$\rm ^{214}Po$}
\def\potwosixteen{$\rm ^{216}Po$}

\def\amtwofortyone{$\rm ^{241}Am$}
\def\fefiftyfive{$\rm ^{55}Fe$}

\def\ms{$\mu$s~}

\def\plusminus{$\pm$~} 

\makeatletter
\AtBeginDocument{\@ifpackageloaded{natbib}{\ifNAT@numbers\if@filesw\immediate\write\@auxout{\string\global\string\NAT@numberstrue}\fi\fi}{}}
\makeatother
\usepackage{amstext}

\begin{document}

\begin{frontmatter}

\title{Tests on NaI(Tl) crystals  for WIMP search at the Yangyang Underground Laboratory  }

\author[snu]{K.W.~Kim}
\author[ibs]{W.G.~Kang}
\author[sju]{S.Y.~Oh}
\author[sju]{P.~Adhikari}
\author[ibs]{J.H.~So}
\author[ibs]{N.Y.~Kim}
\author[ewha]{H.S.~Lee\corref{corres}}
\ead{hyunsulee@ewha.ac.kr}
\author[snu]{S.~Choi}
\author[ewha1]{I.S.~Hahn}
\author[ibs]{E.J.~Jeon}
\author[snu]{H.W.~Joo}
\author[snu]{B.H.~Kim}
\author[knu]{H.J.~Kim}
\author[ibs,sju]{Y.D.~Kim~\corref{corres}}
\ead{ydkim@ibs.re.kr}
\author[ibs,kriss]{Y.H.~Kim}
\author[snu]{J.K.~Lee}
\author[seoul]{D.S.~Leonard}
\author[ibs]{J.~Li}
\author[snu]{S.L.~Olsen}
\author[kriss]{H.S.~Park}

\cortext[corres]{Corresponding authors, Tel:+82-2-3277-3413; fax:+82-2-3277-2372}
\address[ibs] {Center for Underground Physics, Institute for Basic
  Science (IBS), Daejon 305-811, Korea}
\address[snu] {Department of Physics and Astronomy, Seoul National
  University, Seoul 151-747, Korea}
\address[ewha] {Department of Physics, Ewha Womans University, Seoul 120-750, Korea}
\address[ewha1] {Department of Science Education, Ewha Womans University, Seoul 120-750, Korea}
\address[sju] {Department of Physics, Sejong University, Seoul
  143-747, Korea}
\address[knu] {Department of Physics, Kyungpook National University, Daegu 702-701, Korea}
\address[kriss] {Korea Research Institute of Standards and Science, Daejon 205-340, Korea}
\address[seoul] {Department of Physics,  University of Seoul, Seoul 130-743, Korea}

\begin{abstract}
Among the direct searches for WIMP-type dark matter, the DAMA experiment is unique in that
it has consistently reported a positive signal for an annual-modulation signal with a large 
(9.3$\sigma$) statistical significance.  This result is controversial because if it is interpreted
as a signature for WIMP interactions, it conflicts with other direct search experiments that
report  null signals in the regions of parameter space that are allowed by the DAMA observation.
This necessitates an independent verification of the origin of the observed modulation signal
using the same technique as that employed by the DAMA experiment, namely low-background NaI(Tl)
crystal detectors. Here, we report first results of a program of NaI(Tl) crystal measurements
at the Yangyang Underground Laboratory aimed at producing NaI(Tl) crystal detectors with
lower background levels and higher light yields than those used for the DAMA measurements.   
\end{abstract}


\end{frontmatter}


\section{Introduction}
\label{introduction}

Numerous astronomical observations have led to the conclusion that most of
the matter in the Universe is invisible, exotic, and nonrelativistic dark
matter~\cite{Komatsu:2010fb,Ade:2013zuv}.  However, in spite of a considerable amount
of experimental effort, the nature of the dark matter remains unknown.
Weakly interacting massive particles~(WIMPs) are one of the most attractive
candidates for  dark matter particles~\cite{lee77,jungman96}.  In supersymmetric
models for beyond the standard model physics, the lightest supersymmetric particle
(LSP) is a natural candidate for WIMP-type dark matter.  A number of experiments have
made direct searches for a WIMP component of our Galaxy by looking for evidence for
WIMP-nucleus scattering by detecting the recoiling nucleus in ultra-sensitive
low-background detectors~\cite{gaitskell04,baudis12}. 

The DAMA experiment searches for an annual modulation in the detection rate of nuclear
recoils in an array of ultra-low-background NaI(Tl) crystals caused by the Earth's orbital
motion through our Galaxy's dark-matter halo~\cite{bernabei08,bernabei10,bernabei13}.
This experiment, which has been operating for over 15 years, has consistently reported a
positive signal for an annual modulation with a phase that is consistent with expectations
for motion of the Earth relative to the Galactic rest frame. The statistical significance
of the DAMA annual modulation signal has now reacehed $9.3\sigma$.  Other experiments,
including CoGeNT~\cite{aalseth11,Aalseth:2012if,aalseth14}, CRESST~\cite{cresst730kg}, and
CDMS~\cite{agnese13}, have also reported signals that could be interpreted as being possibly
due to WIMP interactions.  However, these signals have marginal significance at the \textless3$\sigma$
confidence level and some inconsistencies with null results using similar detectors that were already reported~\cite{cdex,malbek,cresstnew}.  
Nevertheless, these marginal signals have motivated the same experimental groups and,
in some cases, independent groups to devise experiments to check these signals using similar techniques
but with higher sensitivity.  In contrast, to date, no independent verification of the DAMA
signal in an experiment using the same technique has been done.
 
The DAMA signal, in particular its interpretation as being due to WIMP-nucleus scattering, has 
been the subject of a continuing debate that started with the first DAMA report 15 years ago. 
This is because the WIMP-nucleon cross sections inferred from the DAMA modulation are in conflict
with limits from other experiments that directly measure the total rate of nuclear recoils, such as
XENON100~\cite{aprile12}, LUX~\cite{agnese14}, and SuperCDMS~\cite{akerib14}. 
However, room remains for explaining all of these experimental results without conflict
in terms of nontrivial systematic differences in detector responses~\cite{dsys1,dsys2} and the
commonly used astronomical model for the WIMP distribution~\cite{cmodel}. An unambiguous verification
of the DAMA signal by an independent experiment using similar NaI(Tl) crystals is mandatory.

Reproducing the DAMA measurement requires the independent development of ultra-low-background
NaI(Tl) crystals.  This is because the crystal-growing company\footnote{Crismatec, France} that
supplied the DAMA NaI(Tl) crystals will not produce the same crystals for other experimental
groups.  Recently, the ANAIS group has been developing ultra-low-background NaI(Tl) crystals
with the goal of reproducing the DAMA result~\cite{amare14A,amare14B} and the DM-Ice group reports
background measurements of NaI(Tl) crystals~\cite{dmice}. 
However, to date, no experimental group has produced NaI(Tl) crystals with background levels 
at or below than those used in the DAMA experiment.

The Korea Invisible Mass Search (KIMS) Collaboration has been performing direct WIMP searches
using ultra-low-background CsI(Tl) crystal detectors, which are similar experimental devices to
the NaI(Tl) crystals used in the DAMA experiment. This implementation required
extensive research and development aimed at identifying and reducing internal backgrounds in the
CsI crystals~\cite{kims_pow,kims_crys}.  Null results from 
KIMS reject WIMP-Iodine nuclei interactions as the source of the DAMA signal with very
little model dependence~\cite{hslee07,sckim12}. 
However, because the DAMA results can be interpreted as being primarily due to WIMP-sodium nuclei
interactions, which, for example, would be the case for low-mass WIMPs, it remains necessary
to confirm the DAMA observations with NaI(Tl) crystal detectors. This motivated a program to
develop ultra-low-background NaI(Tl) crystal detectors with lower background levels and and higher
light yields (and, thus, a lower energy threshold) than those of the DAMA experiment in order to
identify unambiguously the origin of the DAMA modulation signature.

\section{Experimental Setup}
To evaluate the NaI(Tl) crystals, we use the experimental setup that was used for the KIMS CsI(Tl)
detector measurements at the Yangyang Underground Laboratory (Y2L), which has a 700~m earth overburden
(2400~m water equivalent).   This includes a 12-module array of CsI(Tl) detectors (total mass of
103.4~kg) inside a shield that consists, from inside out, of 10~cm of copper, 5~cm of polyethylene,
15~cm of lead, and 30~cm of liquid-scintillator-loaded mineral oil to stop external neutrons,
gamma rays, and veto cosmic ray muons.   Each detection module consists of a low-background
CsI(Tl) crystal ($8\times8\times30 ~\rm{cm}^{3}$) with a photomultiplier tube (PMT) mounted at
each end. 

\begin{figure}[!htb]
\begin{center}
\includegraphics[width=0.7\columnwidth]{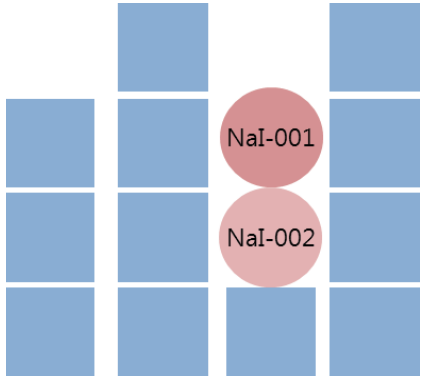}
\end{center}
\caption{Schematic test setup for two NaI(Tl) crystals (circles) with 12 CsI(Tl)
  crystals (squares).}
\label{experimentalsetup}
\end{figure}

Two NaI(Tl) crystals were mounted inside the CsI(Tl) detector array as shown in
Fig.~\ref{experimentalsetup}. In November 2013, we installed the first NaI(Tl) crystal, NaI-001, 
in the CsI array; the second
NaI(Tl) crystal, NaI-002, was added in February 2014, soon after that crystal was delivered and
assembled.

\subsection{NaI(Tl) crystals}
The two NaI(Tl) crystals were produced by the Alpha Spectra
Company.\footnote{http://www.alphaspectra.com} The crystals were grown from NaI powder that
was not of the highest attainable purity; the initial purification was carried out by
Alpha Spectra. The crystals have a cylindrical shape and were cut from a 32~inch ingot
that was grown by the Kyropoulos method.  The detailed sizes of the crystals are listed in
Table~\ref{crystalspec}.  After the crystal surfaces were polished they were wrapped with
a Teflon reflector and inserted into an oxigen-free electronic~(OFE) copper cylinder and encapsulated in a nitrogen
gas environment.  There is a  12.7-mm-thick quartz-plate window at each end of the cylinder,
with optical grease between the crystal and the quartz windows. A 3$''$ PMT is mounted
at each end of the cylinder. 

\begin{table*}[ht]
\begin{center}
\caption{Specifications of the NaI(Tl) crystals used
in this study. The last two columns are the dates the crystals were
grown and transported  to Y2L. ``Transp'' indicates the means by which the
crystals were transported from the United State to Korea.}
\label{crystalspec}
\begin{tabular}{lccccc}\hline
Crystal   & Size                & Mass  & Transp & T (Growth) & T (underground) \\ \hline
NaI-001 & 5$''$(D)$\times$7$''$(L)      &  8.26  &  Air    &  2011.9 & 2013.9 \\
NaI-002 & 4.2$''$(D)$\times$11$''$(L) & 9.2     &  Sea   &  2013.4 & 2014.1 \\\hline
\end{tabular}
\end{center}
\end{table*}

\subsection{Electronics}
Each NaI crystal PMT signal was split and amplified by factors of 30 and 2; the amplified 
signals were digitized by 400- and 64-MHz flash analog-to-digital converters
(FADC), respectively. The corresponding amplification factors for the CsI crystal PMTs
were $\times$100 and $\times$10. The total recorded time window for an event was 40 $\mu$s,
of which 5 \ms is analyzed for the NaI(Tl) crystals and 25 \ms is analyzed for the CsI(Tl)
crystals, reflecting the different decay times of the two materials. 

\begin{figure}[!htb]
\begin{center}
\includegraphics[width=0.7\columnwidth]{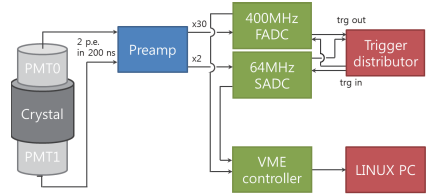}
\end{center}
\caption{Schematic electronics diagram of the KIMS NaI experiment.}
\label{electronics}
\end{figure}

Figure~\ref{electronics} shows a schematic diagram of the detector setup.
 The trigger condition for the CsI(Tl) crystals
is two or more photoelectrons (PEs) in each PMT within a 2-\ms time
window. The NaI(Tl) crystals have a reduced PE requirement of one PE
in each PMTs within a 200 ns window in order to have 
a minimal trigger bias and a low energy threshold. 

The muon rate at Y2L is ($7.0\pm 0.4)\times 10^{-8}/$s/cm$^{2}$/sr~\cite{muon_zhu}. Since the KIMS
muon veto system was under maintenance, no  muon veto is applied to
the data reported here. Approximately 3$\sim$4 muons pass through each
crystal per day.  These induce low-energy events with observed energies
\textless10 keV during a few second time interval after a muon passes through the crystal
because of a long decay-time component of the NaI(Tl) scintillation process. However, most
of these events are rejected by the selection requirements discussed in Section~\ref{sec:cut}
and, so, cosmic ray muon-related events are negligible in the data reported here. 

\section{Signal Calibrations}
The energy calibration of the NaI(Tl) crystals was done with a \amtwofortyone\ 
 source. The detector had one hole of 10~mm in diameter covered by
127-$\mu$m-thick aluminum foil in the center of the encapsulating Cu
container. 
The source was located in front of the hole. The CsI(Tl) crystal calibration
procedure is described in Refs.~\cite{hslee07, sckim12}.

\subsection{Photomultiplier tubes}
Two different types of  low-background PMTs were tested: a metal-packed R11065 and a
glass-packed R12669, both manufactured by Hamamatsu Photonics.\footnote{http://www.hamamatsu.com/}
Table~\ref{photomultipliertubes} shows the specifications
for each PMT.   The R12669 PMT is a modified version of the R6233 Super Bialkali (SBA) PMT
used by DAMA in its recent upgrade~\cite{bernabei12had}. The PMTs we used were
selected for their high quantum efficiency. The radioactivity levels of the
PMTs were measured at underground with a HPGe detector.    
As expected, the metal PMT has a lower radioactivity level than the
glass PMT as listed in Table~\ref{photomultipliertubes}. 
However, the 3$''$ R12669 PMTs  suffer from serious nonlinearity when the signal height
is \textgreater1~V (corresponding to about 1~MeV in energy), and therefore, 
the high-energy region of the NaI(Tl) crystal data taken with these PMTs was not analyzed in this study.

\begin{table*}\caption{Specifications for PMTs tested in this
  study. Radioactivity levels measured with a HPGe detector at
  Y2L. SEL means ``selected for high quantum efficiency.''}
\begin{center}
\begin{tabular}{llcc}
\hline
\multicolumn{2}{c}{PMT}  & $\rm R12669SEL^{b}$   & $\rm R11065SEL^{b}$    \\ \hline
\multicolumn{2}{c}{Photocathode}  & SBA     &  Bialkali \\ 
\multicolumn{2}{c}{Window} & Borosilicate     & Quartz \\ 
\multicolumn{2}{c}{Body} &  Borosilicate    & Kovar \\ 
\multicolumn{2}{c}{Stem} &  Glass    & Glass \\ 
\multicolumn{2}{c}{Gain (HV)} & $1 \times 10^{6}$    & $5 \times 10^{6}$ \\ \hline
Radioactivity$\rm^{a}$     & U(\bitwoforteen ) &  25 \plusminus 5    &  60
\plusminus 5\\ 
(mBq/PMT)       & Th(\actwotwentyeight) &  12 \plusminus 5    & 0.5
\plusminus 0.2\\ 
                       & K(\kforty) &  58 \plusminus 5    & 19
                       \plusminus 2 \\ 
                       & \ratwotwentysix & 60 \plusminus 10 & 5
                       \plusminus 2 \\ 
                       & \tltwooeight & 4 \plusminus 1 &  --- \\ \hline
\end{tabular}

\label{photomultipliertubes}
\end{center}
\end{table*}

\subsection{Single-photoelectron spectra}
\begin{figure}[!htb]
\begin{center}
\includegraphics[width=0.7\columnwidth]{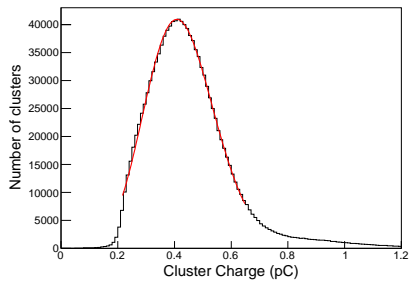}
\end{center}
\caption{The SPE charge distribution measured with the NaI-001 crystal readout by 
an R12669 PMT. This distribution is produced with a \amtwofortyone\ calibration
source in a time window that is 2--5~\ms after the signal start to reduce PE
pileup. 
}
\label{singlephotoelectron}
\end{figure}

For low-energy events, the 400-MHz FADC waveforms were analyzed to
identify clusters of single PEs~(SPEs)~\cite{hslee06}. 
The charge distribution of a SPE, obtained by identifying
isolated clusters at the decay tail of the signal~(2--5~\ms after the signal start) 
in order to suppress multiple PE clusters, is shown in Fig.~\ref{singlephotoelectron}. 

\subsection{Light yields}
\begin{figure}[!htb]
\begin{center}
\includegraphics[width=0.7\columnwidth]{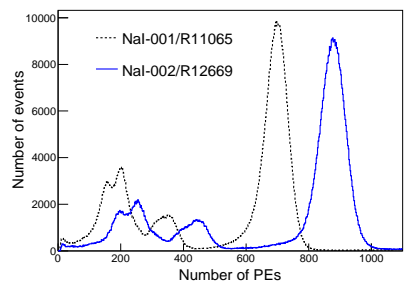}
\end{center}
\caption{The measured number of PEs with the \amtwofortyone\ source calibration for the
NaI-001 and NaI-002 crystals read out by two different PMT types.}
\label{energycalibration}
\end{figure}

Figure \ref{energycalibration} shows the distribution of the number of PEs
obtained using an \amtwofortyone\  source with the NaI-001 and NaI-002 detectors read out
with different types of PMTs. The mean number of PEs detected with the R12669 PMTs
is  22\% greater\ than that detected by the R11065 PMTs.  This is consistent with the ANAIS
test~\cite{amare14A} and reflects the higher quantum efficiency of the SBA photocathode.

\subsection{Scintillation decay time}
\begin{figure}[!htb]
\begin{center}
\includegraphics[width=0.7\columnwidth]{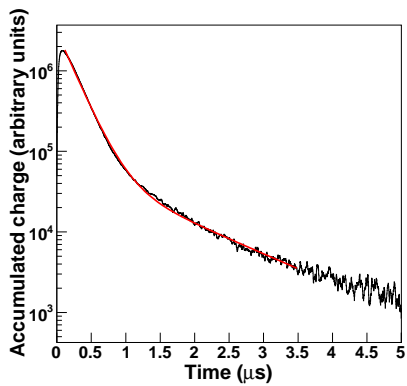}
\end{center}
\caption{Scintillation decay time spectrum of the NaI-001 crystal obtained with 
  \amtwofortyone\ source data and fitted with two exponential functions.
 }
\label{decaytime}
\end{figure}

We study the scintillation decay time of the NaI(Tl) crystals from the time
response to signals induced by a radioactive $\gamma$-ray source. Figure~\ref{decaytime}
shows the integrated signal shapes for 59.54~keV $\gamma$-rays
from a \amtwofortyone\  source. The data are fitted with two
exponentials with time constants of 0.22 and 1.17 $\mu$s. The fast component
accounts for  $\sim$83\% of the total light yield.

\section{Natural background}
To produce ultra-low-background crystals, contamination from internal natural
radioisotopes has to be reduced. Table \ref{internalbackgrounds} shows the
measured results of the internal backgrounds for the two crystals.  In this
section we describe some details of these measurements.

\begin{table}\caption{Backgrounds from the internal radioactive contaminants in the
  NaI(Tl) crystals. The units for all the values are  mBq/kg. For ``Total alphas,'' each
  alpha particle is counted as one decay.}

\begin{center}
\begin{tabular}{lccccc}
\hline Radionuclei                          & NaI-001              &  NaI-002  \\ \hline
\utwothirtyeight\ (by \bitwoforteen)  & \textless 0.0003     &  \textless 0.0015 \\
\thtwotwentyeight\ (by \potwosixteen) &  \textless 0.013     &  0.002 \plusminus 0.001\\
\kforty                              & 1.25 \plusminus 0.09 & 1.49 \plusminus 0.07 \\
\pbtwoten                            &  3.28 \plusminus 0.01     & 1.76 \plusminus 0.01\\
Total alphas                         &  3.29 \plusminus 0.01    &  1.77 \plusminus 0.01  \\\hline
\end{tabular}
\label{internalbackgrounds}
\end{center}
\end{table}

\subsection{\kforty\  background} The most serious internal background 
contamination is \kforty\ because of the low-energy x-ray that is produced during its
electron capture decay process, which proceeds via a transition 
to an excited state of \arforty\ with a branching ratio of 10\%. 
This decay generates a $\sim$3~keV x-ray in coincidence with a 1460~keV $\gamma$-ray.
If the accompanying 1460~keV $\gamma$-ray escapes from the crystal, the event consists
of a single 3~keV hit.  The \kforty\ level in the DAMA crystals is in the 10--20~ppb
range~\cite{bernabei08had}. 

\begin{figure*}[!htb]
\begin{center}
\begin{tabular}{cc}
\includegraphics[width=0.48\textwidth]{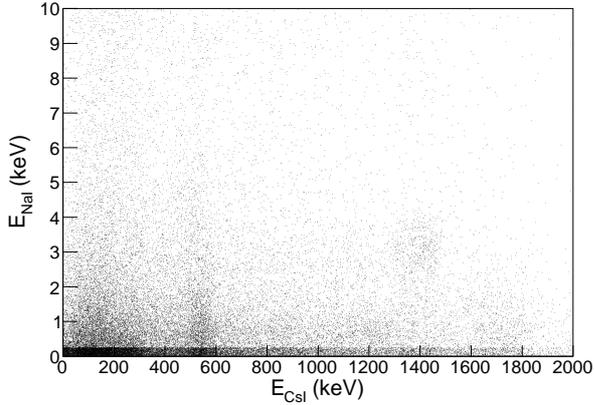}&
\includegraphics[width=0.48\textwidth]{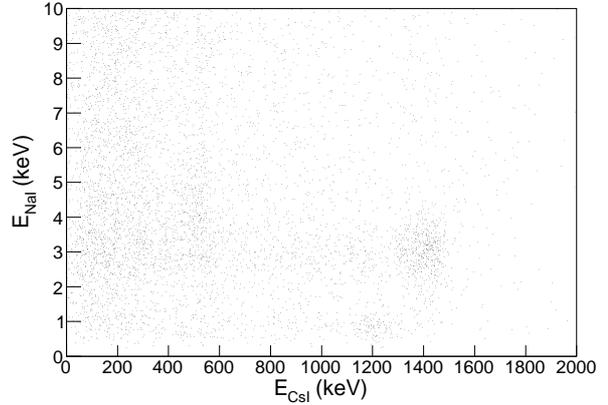}\\
 (a) No cut  & (b) After PMT noise cuts  \\
\end{tabular}
\end{center}
\caption{Scatter plots of energy in a NaI(Tl) crystal~($E_{\rm NaI}$) {\it versus}
  that in a surrounding CsI(Tl) crystal~($E_{\rm CsI}$) before (a) and after (b) the application
  of event selection requirements that remove PMT noise events.  The cluster at 3 keV is due to
  \kforty\ and that at 1 keV is due to $^{22}$Na.
  }
\label{csi_nai_2d}
\end{figure*}

We studied coincidence signals between NaI~($\sim$3 keV) and CsI~(1460~keV) detectors to identify
unambiguous \kforty\ decays.  Figure~\ref{csi_nai_2d}(a) shows a scatterplot of NaI {\it versus} CsI
detector reponse without the application of any of the requirements used to remove PMT noise-induced
signals that are discussed in Section~\ref{sec:cut} below;
Fig. \ref{csi_nai_2d}(b) shows the same plot after the application of these requirements.
In Fig.~\ref{csi_nai_2d}(b), the event cluster near 3~keV in NaI and 1460~keV
in a surrounding CsI crystal is from \kforty\ decays.  Another, smaller cluster near  $\sim$1 keV in the NaI and
1270 keV in one of the CsI crystals is from \natwentytwo decays. 
By comparing the rate for \kforty\ induced events with a Geant4-based detector
simulation~\cite{geant4} for the CsI and NaI crystal setup, we determine the \kforty\
 contamination in the NaI crystals to be  $41.4\pm3.0$~ppb
($1.25\pm0.09$ mBq/kg) and $49.3\pm2.4$~ppb ($1.49\pm0.07$~mBq/kg), 
respectively, for NaI-001 and NaI-002. 
This is close to the $^{40}$K levels measured in the ANAIS-25 crystals~\cite{amare14A,anaisk40}.


\subsection{\utwothirtyeight\  background}

\begin{figure}[!htb]
\begin{center}
\includegraphics[width=0.7\columnwidth]{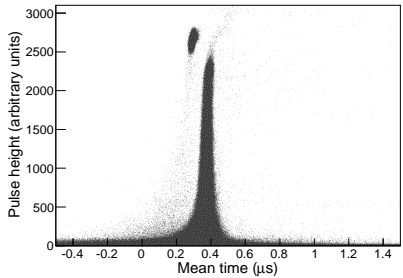}
\end{center}
\caption{A scatter plot of maximum height {\it versus} mean time of event signals from NaI-001. Alpha-induced events
are well separated from $\gamma$-induced signals because of their shorter decay time.  }
\label{r11065alphameantime}
\end{figure}

Although the PMTs are nonlinear at high light output,  alpha-induced events
inside the crystal can be identified by the mean time of
the signal, defined as
\begin{equation} 
\langle t\rangle \equiv \frac{\sum^{}_{i} A_{i}t_{i}  }{\sum^{}_{i} A_{i} }.
\end{equation} 
Here $A_{i}$ and $t_{i}$ are the charge and time of each cluster
(for low energies) or digitized bin (for high energies).
Figure~\ref{r11065alphameantime} shows a scatter plot of the pulse height {\it versus} mean time
for event signals from NaI-001. Alpha-induced events are clearly separated from gamma-induced events
in the high-energy region because of the faster decay times of alpha-induced signals.

\begin{figure}[!htb]
\begin{center}
\includegraphics[width=0.7\columnwidth]{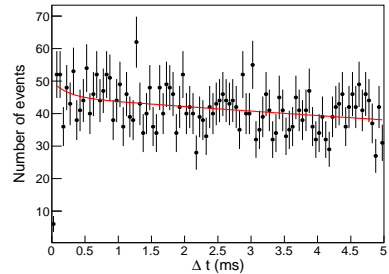}
\end{center}
\caption{The $\beta$--$\alpha$ coincidence time spectrum from the NaI-001 crystal.  
There is no significant exponential decay from a \bitwoforteen\ $\beta$  and \potwoforteen\  $\alpha$-decay component and an upper
  limit is determined from the fit.  }
\label{alphadeltat}
\end{figure}

Because of the nonlinearity of high-energy signals, alpha particles from different nuclides
cannot be distinguished event-by-event by their measured energies.  Instead we determine the level of
\utwothirtyeight\ chain contaminants by exploiting the \potwoforteen\  237~$\mu$s mean lifetime that occurs
between \bitwoforteen\ $\beta$-decay  and \potwoforteen\  $\alpha$-decay, 
a technique that was used successfully for contamination measurements in the KIMS CsI
crystals~\cite{kims_crys}.
Figure~\ref{alphadeltat} shows the distribution of measured time intervals between an
alpha-induced event and its immediately preceding event.  Since there is no significant
exponential component observed with the 237-\ms decay time of $^{214}$Po, an upper
limit on the activity level is determined.   This analysis shows that the contamination levels from the 
\utwothirtyeight\ chain are already sufficiently low for a WIMP dark matter search. 

\subsection{\thtwothirtytwo\ background}
\begin{figure}[!htb]
\begin{center}
\includegraphics[width=0.7\columnwidth]{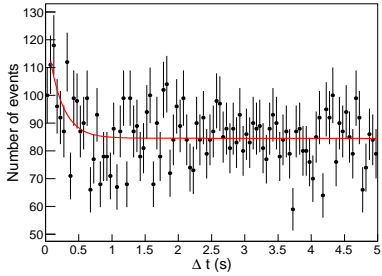}
\end{center}
\caption{The time difference distribution between two alpha events in the NaI-001 crystal. The
exponential component below 1~s is due to the sequential decays of
\rntwotwenty\ and \potwosixteen .
}
\label{alphaalphadeltat}
\end{figure}

Contamination from the \thtwothirtytwo\ chain is studied by using $\alpha\text{--}\alpha$ 
time interval measurements in the crystals. In this case we look for
a \potwosixteen\ $\alpha$-decay component with a mean time of 209~ms
following its production via \rntwotwenty~$\rightarrow $~\potwosixteen\ $\alpha$-decay.  
Figure \ref{alphaalphadeltat} shows the distribution
of the time difference between two alpha events. 
As one can see in Fig.~\ref{alphaalphadeltat}, there is a small
exponential component with the  \potwosixteen\ decay time in NaI-001; 
there is no such signal in NaI-002 and we set an upper limit for the contamination in that crystal. 
These $\alpha\text{--}\alpha$ event rates can translate into contamination levels 
from the \thtwotwentyeight\ series in the \thtwothirtytwo\ chain.
The \thtwothirtytwo\ contamination levels of the two crystals, listed in
Table~\ref{internalbackgrounds}, are also sufficiently low.  

\subsection{\pbtwoten\ background}
The levels of 
\utwothirtyeight\ and \thtwothirtytwo \ contamination measured in both crystals are
too low to account for the total observed alpha particle rate, which suggests that they are due to
decays of \potwoten\ nuclei that originate from \rntwotwotwo\ contamination that occured sometime during the
powder and/or crystal processing stages.  This is confirmed by the observation of a 46-keV $\gamma$
peak that is characteristic of \pbtwoten. 

\begin{figure*}[!htb]
\begin{center}
\begin{tabular}{cc}
\includegraphics[width=0.48\textwidth]{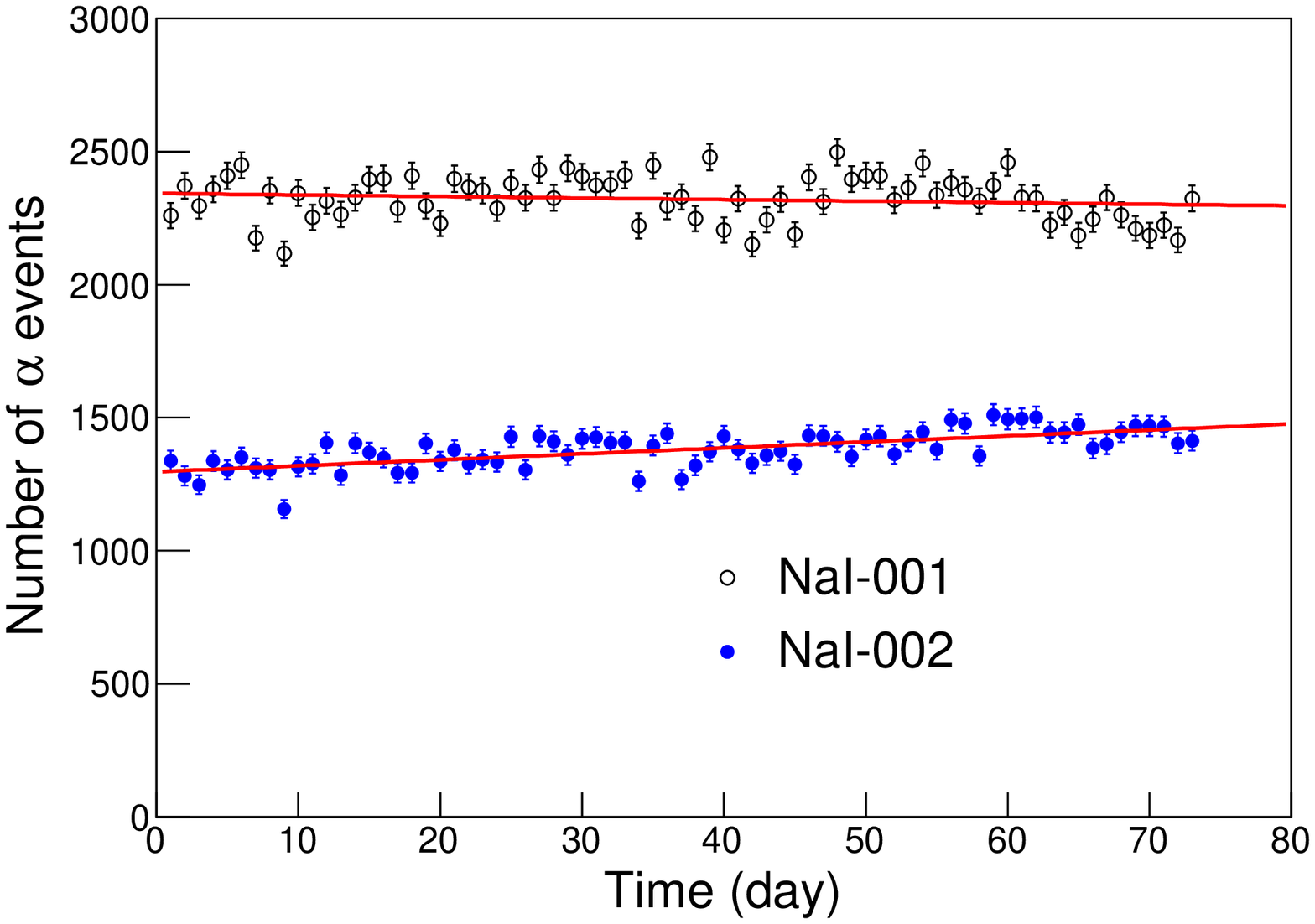}&
\includegraphics[width=0.48\textwidth]{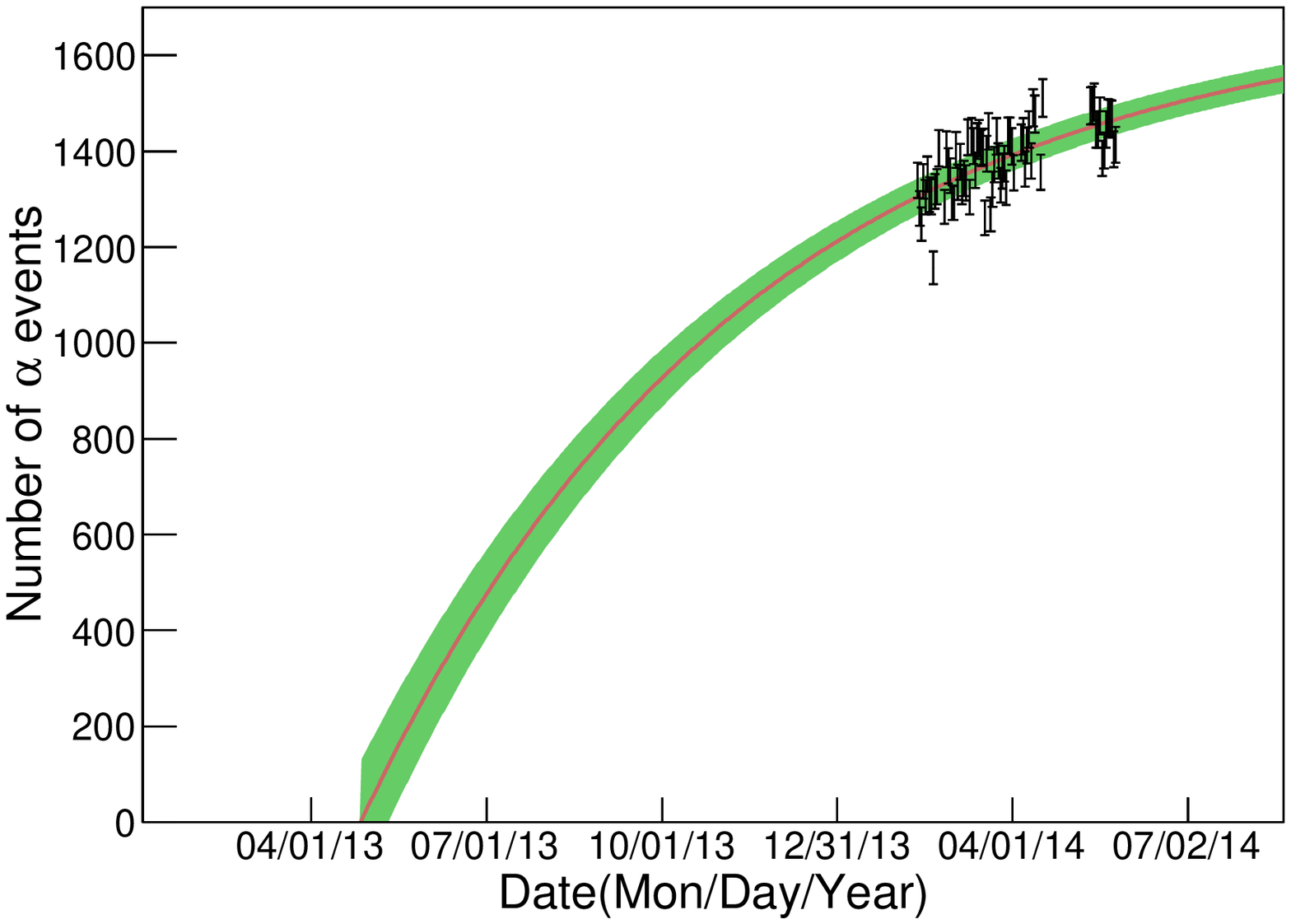}\\
(a)  & (b)  \\
\end{tabular}
\end{center}
\caption{(a) Total number of alpha particles per day for the two NaI
  detectors. (b) The alpha activity increase in NaI-002 is 
  fitted with a model in which a nearly instantaneous \pbtwoten\
  contamination is assumed.
}
\label{alpharatesdaqtime}
\end{figure*}

The time change in the total alpha rate provides information about when the \rntwotwotwo
contamination occurred. After \rntwotwotwo contamination, the number of \pbtwoten\
nuclei increases as does the \potwoten\ $\alpha$-decay rate. After about three years,
equilibrium is reached and the \potwoten\ activity becomes constant. Figure~\ref{alpharatesdaqtime} 
shows the total alpha rates in the two crystals as a function of data-taking time (days).  
The NaI-001 and NaI-002 crystals
emit about 2344 and 1334 alpha particles per day, respectively. After considering the
crystal masses, we find that the NaI-002 alpha activity is less than that for
NaI-001 by almost a factor of two.  Moreover, we also see that the NaI-002 crystal's
alpha activity is increasing with time. 

For \potwoten , the alpha activity will increase as 
\begin{equation}  
R_{\alpha}(t)\approx A(1-e^{-(t-t_{0})/ \tau_{\text{Po}}}),
\end{equation}  
where $\tau_{\text{Po}}$ is the decay time of \potwoten\ (200 days) and $t_{0}$ is the
time the initial \pbtwoten\ contamination occurred, assuming that the contamination 
happened suddenly. The measured alpha rate was fitted to this equation,
and the results indicate that the contamination occurred at the end of April, 2013.
This coincides with the time that the crystal was grown, and we conclude that the
contamination occurred then. 

\section{Background from cosmic excitation}
\begin{figure*}[!htb]
\begin{center}
\begin{tabular}{cc}
\includegraphics[width=0.48\textwidth]{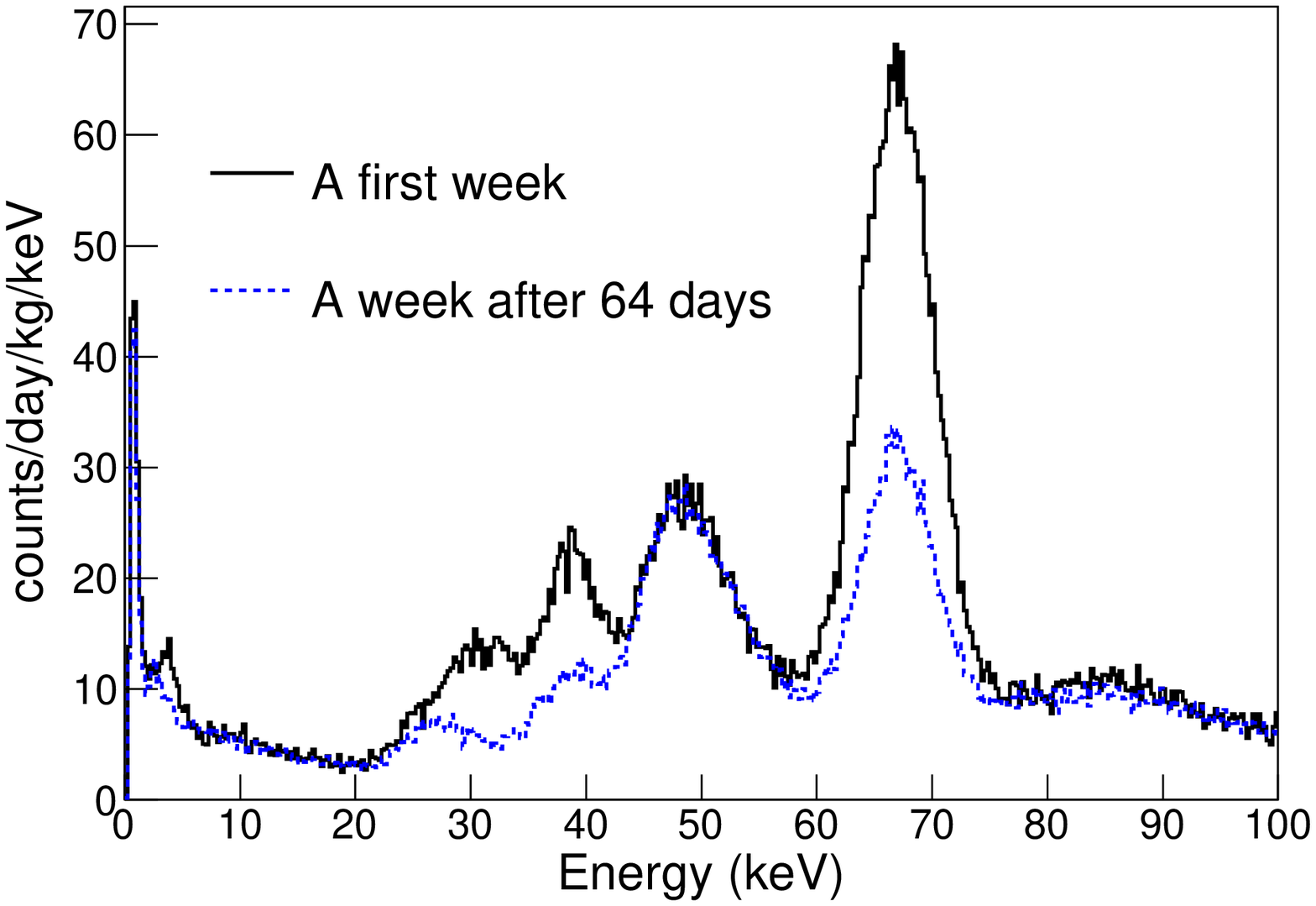}&
\includegraphics[width=0.48\textwidth]{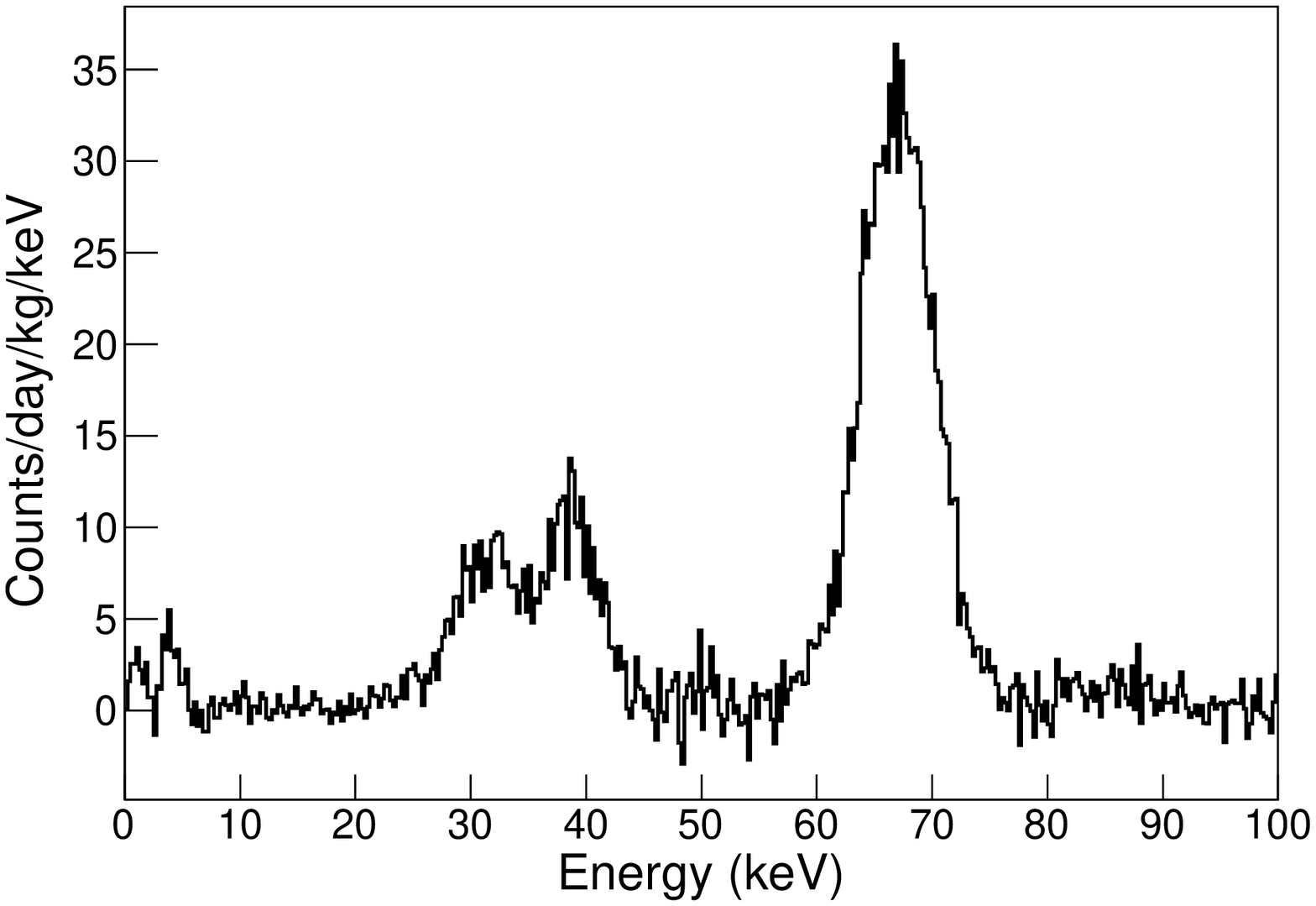}\\
(a)  & (b)  \\
\end{tabular}
\end{center}
\caption{(a) Energy spectra of the NaI-001 crystal during the first week after the
arrival of the crystal underground and for a
  week following a 64 day interval. The peak around 50-keV is due to
  \pbtwoten\ decay inside the crystal. (b) The difference between the initial
and the delayed energy spectra, which shows short life-time cosmogenic activations.  
}
\label{cosmogenic_lastfirst}
\end{figure*}

The two crystals were transported from the U.S. to Korea by different means,
NaI-001 by air and NaI-002 by sea, in order
to understand the cosmogenic-activation-dependence on the delivery method. 
Figure~\ref{cosmogenic_lastfirst}(a) shows the energy spectra for the NaI-001 crystal
during the first week and for a week-long period after a 64-day delay after the arrival
of the crystal underground.  Figure~\ref{cosmogenic_lastfirst}(b) shows the difference
between the first and second measurements.  The peak at  68.7-keV is the sum of 
$\gamma$-rays and x-rays from \ionetwentyfive electron capture decay. The lower energy peak
also can be identified as originating from iodine and  tellurium decays. We found
significantly lower cosmogenic activation in the NaI-002 crystal, and we
conclude that surface transportation is mandatory for low-background crystals. 

\natwentytwo can be produced through the ($n,2n$) reaction on
\natwentythree by energetic cosmic neutrons at sea level. It decays 
via positron emission (90\%) and electron capture (10\%) followed by 
1270-keV gamma  emission with a mean lifetime of 3.8~years. The electron capture decay produces $\sim$0.8~keV
x-rays. Therefore about 10\% of the \natwentytwo decay will produce 0.8-keV x-rays
and 1270-keV $\gamma$-rays at the same time. The $\gamma$-$\gamma$ coincidences
show up in Fig.~\ref{csi_nai_2d} as a cluster of events
below the  \kforty\ decay events. The more frequent $\beta^ +$ decay channel does not generate
low-energy x-rays. The 0.8-keV events are useful for studying selection efficiencies in
the 1-keV energy region.


\section{PMT noise background}
\label{sec:cut}
Photomultiplier tubes are known to generate low-energy noise signals primarily via four
different mechanisms.  First, radioactive decays of U, Th, and K inside the PMT materials
generate ultraviolet and/or visible photons directly inside the PMTs.
Second, the high voltage applied to the PMT can cause charge to accumulate 
somewhere in the PMT and subsequently discharge, producing a flash. Third, PMT
dark currents will produce accidental coincidences between two PMTs that satisfy
the trigger condition. Fourth, large pulses can result from afterpulsing 
produced by ionized residual gas inside the PMT.   In fact, PMT noise involves complex
phenomena that are far from being completely understood.

\subsection{Accidental background from dark current}
All PMTs normally have some level of dark current; this is essentially due to 
SPEs that are spontaneously emitted from the photocathode. In the PMTs used
in these measurements, the SPE rates vary from PMT to PMT, and are typically 
of order $\sim$1 kHz.   The accidental rate of two dark-current SPEs from two PMTs
within 200~ns is $\sim$0.2~Hz. These accidental events generally have a
very low charge but, because of the large variation in the charge registered
for a SPE, the energy could occasionally be \textgreater1~keV. Therefore,
we require at least two PEs in both PMTs to reject events from
dark-current-induced accidentals.

\subsection{DAMA cut}
\begin{figure*}[!htb]
\begin{center}
\begin{tabular}{cc}
\includegraphics[width=0.48\textwidth]{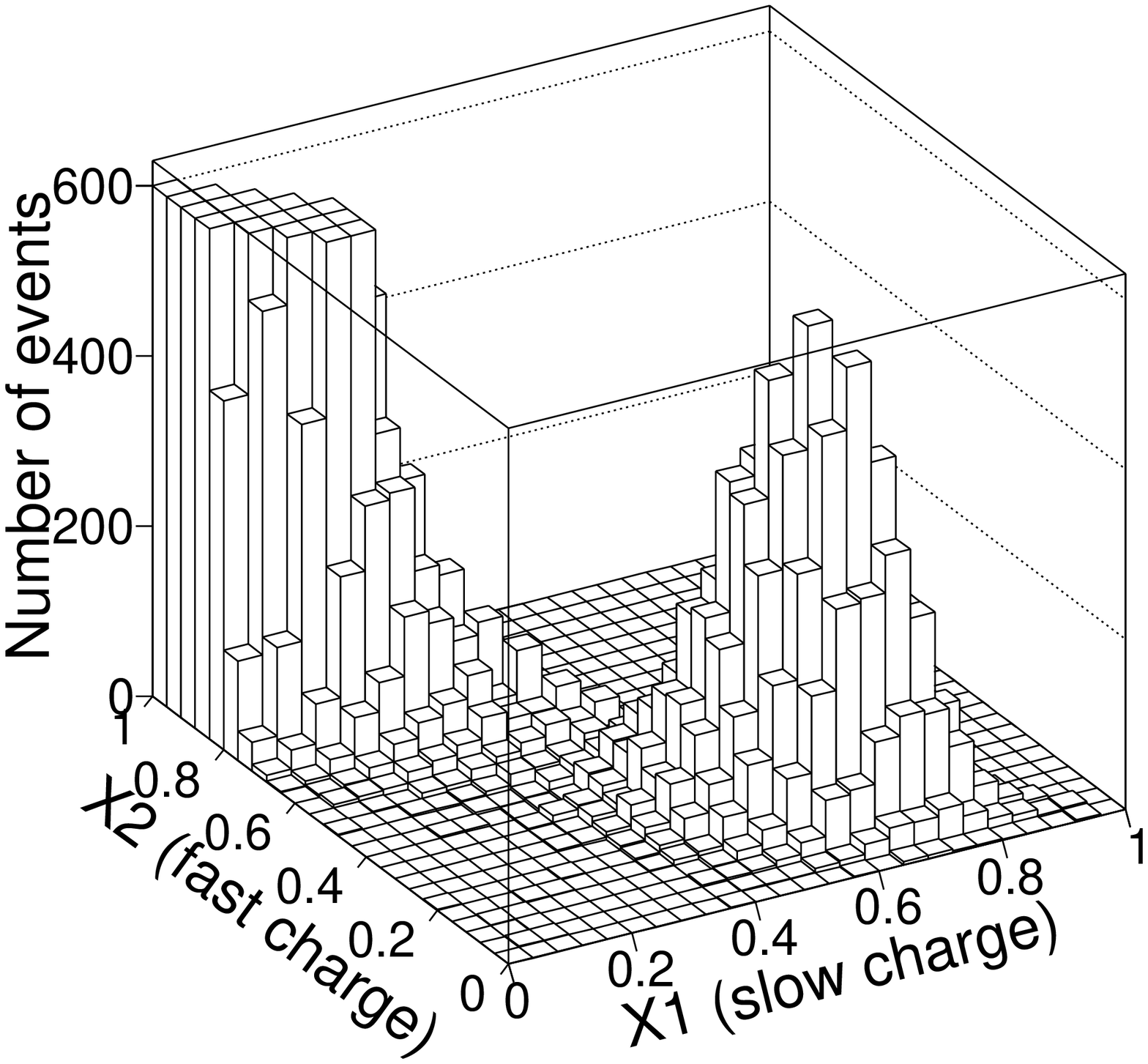}&
\includegraphics[width=0.48\textwidth]{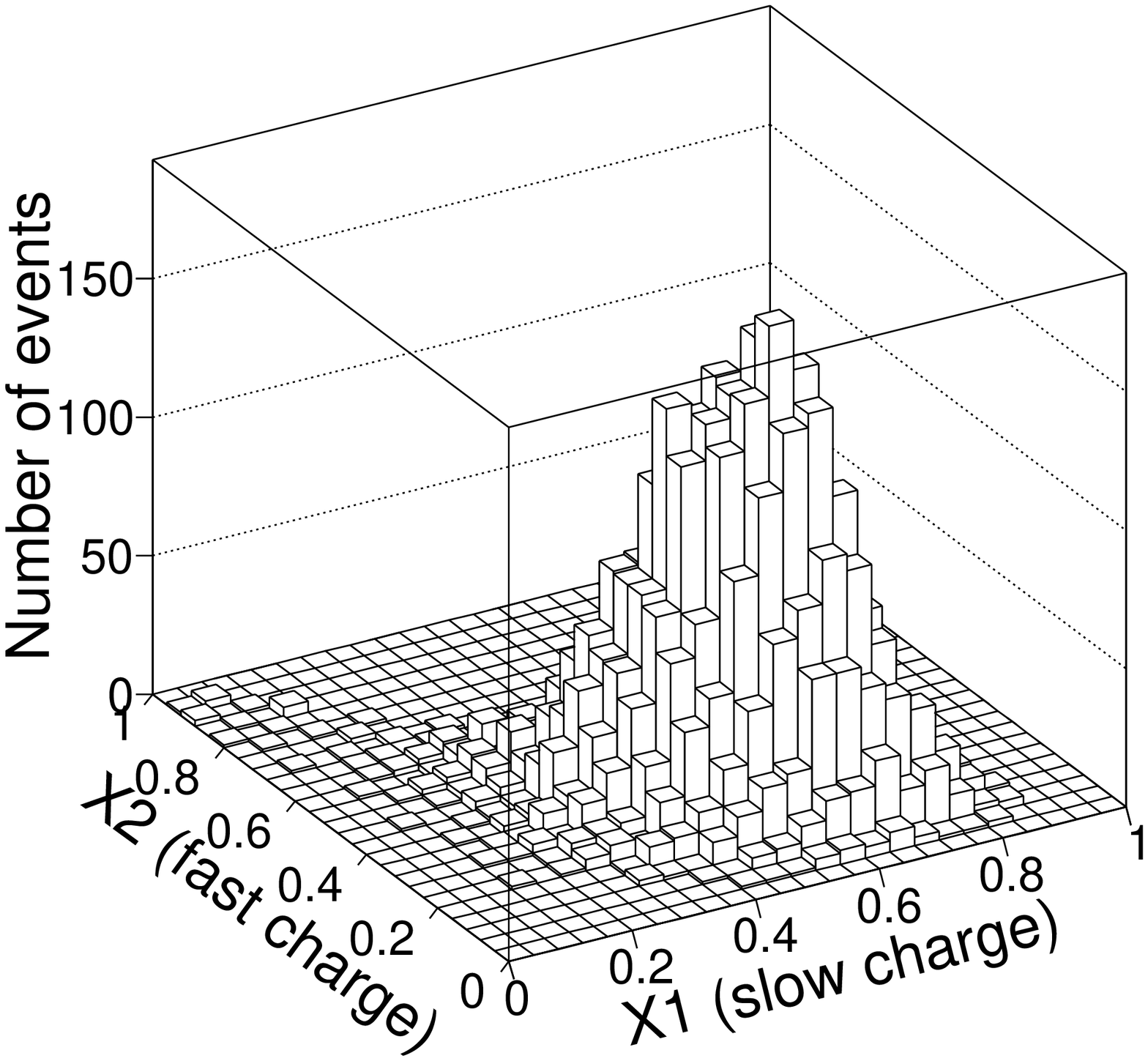}\\
(a) Background data  & (b)\fefiftyfive\ calibration data  \\
\end{tabular}
\end{center}
\caption{Two-dimensional plots of ``fast'' and ``slow''
  charges for background~(a) and low-energy x-ray source calibration~(b) data. 
  PMT noise events have larger ``fast'' charges and signal events
  have larger ``slow'' charges. }
\label{damacutlego}
\end{figure*}

The DAMA group reported a signal selection criteria for efficiently removing the PMT noise
events from their NaI(Tl) detectors that exploits the fact that noise pulses
are generally fast.  The DAMA requirement places restrictions on the ratio of 
``fast'' charge (0--50~ns), X1, and ``slow'' charge (100--600~ns), 
X2~\cite{bernabei12had,bernabei08had}. We examined the DAMA parameters for
our NaI(Tl) crystals. Figure~\ref{damacutlego} shows a two-dimensional X2
{\it versus} X1 scatter-plot for events in the 2--4~keV energy range both for background
data (a) and for data taken with an \fefiftyfive\ source (b).  The figure shows that our
discrimination between noise and signal is very efficient, similar to the DAMA
results~\cite{bernabei08had}.  
A rejection rate of the 2--4~keV WIMP search data is approximately 84\%, however approximately 86\% 
of the $\sim$3~keV $^{40}$K coincidence events is remained. 

\subsection{Asymmetry cut}
Although a large fraction of PMT noise events below 5~keV are removed by the DAMA
requirement, we find that some PMT noiselike events remain.  We, therefore, developed further
requirements to remove these events. We define the asymmetry between the two PMT signals as 
\begin{equation}
\text{asymmetry} \equiv \frac{Q_{1}-Q_{2}}{Q_{1}+Q_{2}},
\end{equation}
where $Q_{1}$ and $Q_2$ are the charges measured by the two PMTs. This asymmetry allows us
to locate where the event occurred inside the crystal. 
A similar study of asymmetry cuts with NaI crystals was reported previously~\cite{naiad}.

\begin{figure*}[!htb]
\begin{center}
\begin{tabular}{cc}
\includegraphics[width=0.48\textwidth]{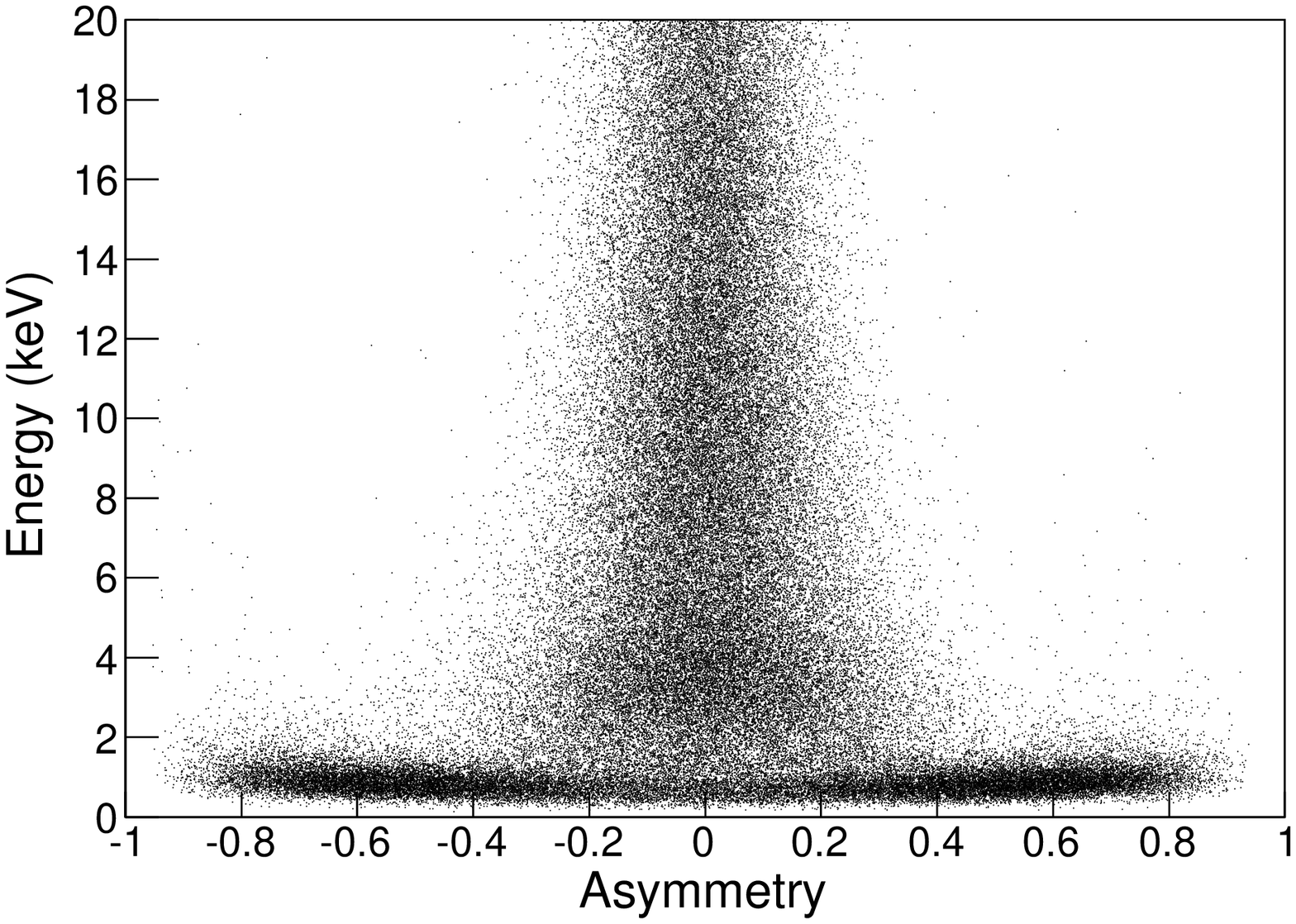}&
\includegraphics[width=0.48\textwidth]{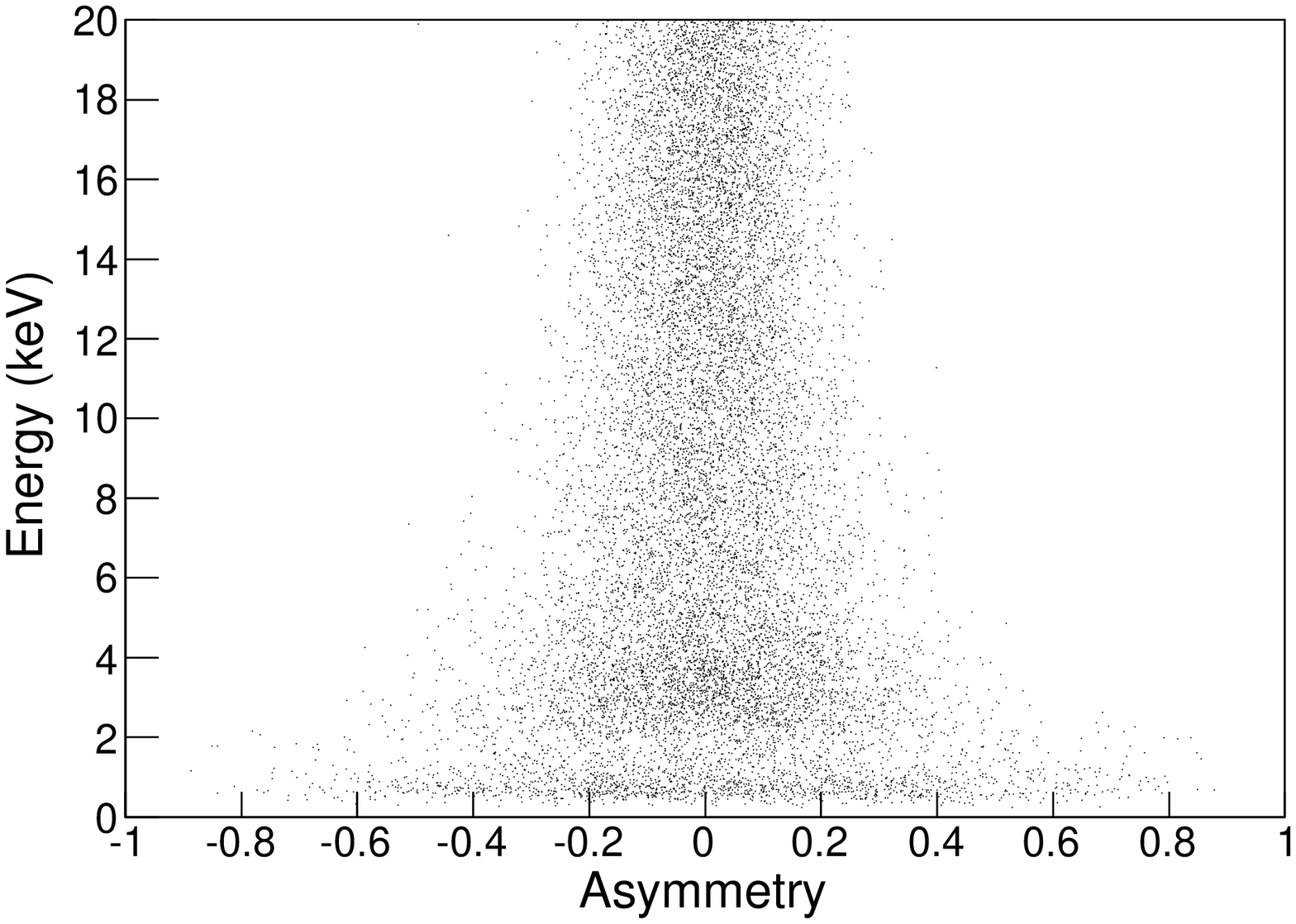}\\
(a) Single-hit events  & (b) Multiple-hit events  \\
\end{tabular}
\end{center}
\caption{Asymmetry {\it versus} energy plots of the NaI-001 coupled with an R12669
  PMT for single-hit events (a) and multiple-hit events (b).
  Many single-hit events with energy below 3~keV have large asymmetries that are
  attributed to PMT noise. 
}
\label{asymmetrycut}
\end{figure*}

To characterize PMT noiselike events, we used multiple hit events in which
two or more detector modules satisfy the trigger condition. 
Figure~\ref{asymmetrycut} shows a two-dimensional asymmetry {\it versus} energy scatter plot
for single-hit~(a) and multiple-hit~(b) events. 
These data are obtained from the NaI-001 crystal coupled with R12669 PMTs. In the data shown
in these figures, the DAMA requirement has already been applied. 
The multiple-hit events have $|{\rm asymmetry}|<0.6$, while many single-hit events with
energy below 3~keV have asymmetries that are  even larger than those for real events that
occur near the edge of the crystal.  This suggests that these events are caused by visible
light produced near one of the PMTs.  In contrast, the NaI-002 detector with the R11065
metal PMTs do not show many events with such large asymmetries.  
Events with $|{\rm asymmetry}|>0.6$ are rejected.

\subsection{Cluster charge cut}
\begin{figure}[!htb]
\begin{center}
\includegraphics[width=0.7\columnwidth]{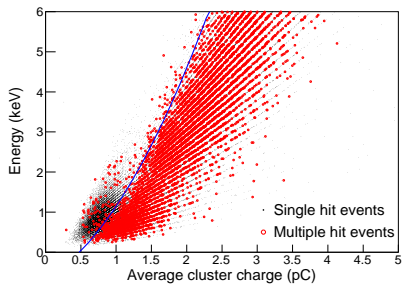}
\end{center}
\caption{Energy versus the average charge of clusters for single-hit events~(dots) and
multiple-hit events~(open circles). 
There are additional noise events for the single-hit events in lower cluster charge
regions. The solid line shows the cut condition to remove such noise events. }
\label{qcnccut}
\end{figure}

We found a peculiar class of low energy events in the single hit sample that are made
up of SPE clusters that are spread nearly uniformly over a few hundred nanosecond interval.
These are evident in the scatter-plot of the total energy {\it versus} the average charge of
the energy clusters (total charge/number of clusters) shown in  Fig.~\ref{qcnccut},
where the black dot entries are for single-hit events and the red circles are for multiple-hit events.
These data were obtained with the NaI-001 crystal coupled to the glass R12669 PMTs, and 
the DAMA and asymmetry selection requirements were applied. 
A distinct cluster of low energy signals with an average cluster charge consistent with that
for a SPE shows up for single-hit events.  While the source of these events is still not
understood, since they do not show up in multiple hit events, they are considered likely to
be induced by PMT-noise. Although this phenomenon requires additional study, 
for now we veto events that lie to the left of the solid curve shown in the figure.

\section{Background model}
\begin{figure}[!htb]
\begin{center}
\includegraphics[width=0.7\columnwidth]{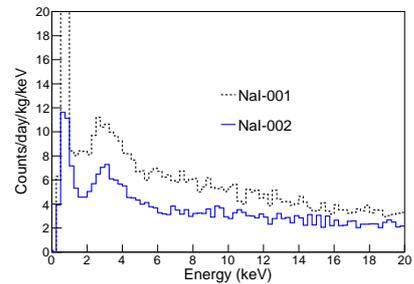}
\end{center}
\caption{Background levels in the two crystals after the application of the PMT noise
  rejection requirements. Here we show data obtained with the R12669 glass 
  PMTs.
}
\label{backgroundlevel}
\end{figure}

Figure~\ref{backgroundlevel} shows the background levels of the two crystals
coupled to R12669 PMTs after the application of all of the event selection criteria
discussed before. NaI-002 has a much lower background level than that of NaI-001,
because of its lower cosmogenic activation as a result of its surface delivery and
its lower \pbtwoten\ contamination. Its  background level at 6~keV is
$\sim$3~counts/kg/keV/day. The figure also shows that the new PMTs with higher quantum
efficiency may enable us to lower the energy threshold to near 1~keV if we have an additional
understanding of the PMT noise below 2~keV.

\subsection{Background simulations}

\begin{figure*}[!htb]
\begin{center}
\begin{tabular}{cc}
\includegraphics[width=0.48\textwidth]{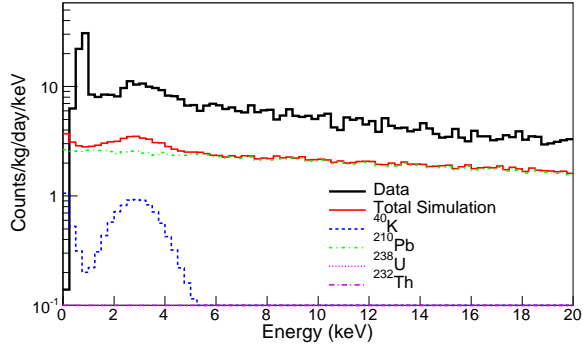}&
\includegraphics[width=0.48\textwidth]{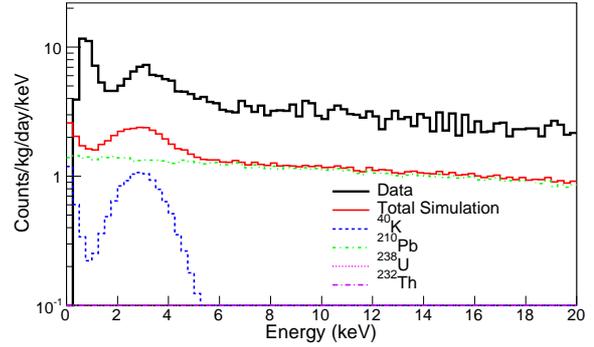}\\
(a) NaI-001 & (b) NaI-002 \\
\end{tabular}
\end{center}
\caption{The measured background levels in the NaI crystals compared with
  simulations of backgrounds from internal contamination in the crystals.  
}
\label{backgroundsimulation}
\end{figure*}

We simulated the background spectra with Geant4-based detector simulations of \pbtwoten,
\utwothirtyeight, \thtwothirtytwo, and \kforty\ with contamination levels set at the measured
values for each crystal. Figure~\ref{backgroundsimulation}
shows the data and the simulated spectra for the two crystals. For the current NaI(Tl) crystals,
the significant remaining backgrounds are from \pbtwoten, \kforty, and 
PMT noise. The U, Th, and Ra internal contamination levels in the crystal produce backgrounds
at low energies that are already sufficiently small: {\it i.e.} \textless0.1 counts/keV/kg/day.
The measured background levels of NaI crystals for energies below 10~keV is higher than that of the 
internal background simulation by $\sim$2~counts/keV/kg/day.  This difference is attributed to
$\gamma$-rays from sources that are exterior to the crystal, {\it i.e.}, the PMTs, the surrounding CsI
crystals, and the materials of the surrounding shield.  In addition to these constant backgrounds,
there are also  \ionetwentyfive and \teonetwentyfive cosmogenic backgrounds that are continuously
decreasing as a function of time. Further simulations will clarify and quantify the contributions
from each external source as well as the contributions from cosmogenic activations to the total
background level.

\section{Perspectives}
The DAMA experiment has been consistently showing a significant annual modulation with two
different experimental arrangements that has persisted over the past 15 years. To check these
results, it is necessary to use the same target material and preferable to have a lower
threshold and reduced background levels.  Achieving  a software energy threshold of 1 keV seems
feasible because of the high crystal light output and the high quantum efficiency of the new
PMTs. Background levels can be significantly improved from the current measured background
level of  $\sim$3~counts/keV/kg/day~(at 6~keV) by growing the crystals from purer NaI powder and
in a Rn-free environment.   Because the main internal backgrounds are due to \pbtwoten\ and
\kforty, we are attempting to reduce these contaminations below the background level of
0.2~counts/keV/kg/day for each source, and the prospects of starting with
purer powder from Sigma Aldrich\footnote{http://www.sigmaaldrich.com} are promising. 

In addition to the internal backgrounds, the external backgrounds
need to be controlled well below 0.5~counts/keV/kg/day. 
Low-background, metal-housed PMTs with lower radioactivity specifications
are commercially available and it is possible to use high efficiency SBA
photocathodes with these tubes. We are  working closely with the Hamamatsu Company
to develop a PMT that is better suited for a low-background NaI(Tl) crystal detector module.  

Further, we expect a significant reduction in the internal or external backgrounds 
by the immersion of the NaI(Tl) crystal array inside a liquid scintillator box that
provides an active veto capability. A naive simulation shows that  $\sim$70\% of the
PMT-initiated backgrounds below 10 keV can be vetoed. A performance test with a single
NaI crystal is in progress. 

\section{Conclusion}
We tested the performance of two large NaI(Tl) crystals as part of a program to
develop ultra-low-background NaI crystals for  WIMP searches. We developed
selection requirements that are effective for reducing PMT-noise induced background signals.  
Based on this effort, we achieved a background level of $\sim$3~counts/keV/kg/day at 6~keV and
a \textless2~keV energy threshold.  A number of efforts are being pursued that are aimed at
further reduction of these  backgrounds. The successful development of ultra-low-background and
low-energy-threshold NaI crystals with much reduced PMT background will guarantee  
a definitive and unambiguous test of the DAMA experiment's annual modulation signal. 

\section*{Acknowledgments}
We thank the Korea Hydro and Nuclear Power (KHNP) company for
providing the underground laboratory space at Yangyang. This research
was funded by Grant No. IBS-R016-D1 and was supported by the Basic
Science Research Program through the National Research Foundation of Korea
(NRF) funded by the Ministry of Education~(NRF-2011-35B-C00007).

\end{document}